\documentclass[preprint,amsmath,amssymb,aps,showkeys,showpacs]{revtex4}
\usepackage[english]{babel}
\usepackage{graphicx}
\usepackage{graphics}
\usepackage{amsmath}
\usepackage{dcolumn}
\usepackage{amssymb}
\usepackage{bm}

\begin{document}
\title{Doubly anharmonic oscillator under the topological effects of a screw dislocation}
\author{Knut Bakke}
\email{kbakke@fisica.ufpb.br}
\affiliation{Departamento de F\'isica, Universidade Federal da Para\'iba, Caixa Postal 5008, 58051-900, Jo\~ao Pessoa-PB, Brazil.}

\begin{abstract}

We consider an elastic medium with the distortion of a circular curve into a vertical spiral, and investigate the influence of this topological defect on the doubly anharmonic oscillator. We show that the Schr\"odinger equation for the doubly anharmonic oscillator in the presence of this linear topological defect can be solved analytically. We also obtain the exact expressions for the permitted energies of the ground state of the doubly anharmonic oscillator, and show that the topology of the screw dislocation modifies the spectrum of energy of the doubly anharmonic oscillator.

\end{abstract}


\maketitle

\section{Introduction}

A great interest in the doubly anharmonic oscillator is in field theory \cite{doub8,doub9,doub10,doub11,doub,doub13,doub1}. In these studies, the doubly anharmonic oscillator has been dealt with the perturbation theory. This way of treating this problem has inspired several works that focus on the analytical solutions to the Schr\"odinger equation for the doubly anharmonic oscillator \cite{doub2,doub3,doub4,doub5,doub7}. From this perspective, an interesting point of discussion is the behaviour of the doubly anharmonic oscillator in the presence of torsion, since torsion has a great interest in quantum field theory \cite{torsion1,torsion2,torsion3}, condensed matter physics \cite{kleinert,kat,val} and gravitation \cite{put,valdir,bf2,vb}. In particular, torsion effects in an elastic medium are associated with the presence of topological defects in a crystal. An example is the distortion of a circular curve into a vertical spiral, which is called as a screw dislocation \cite{val,put}. In recent decades, several authors have discussed the influence of this screw dislocation on quantum systems \cite{fil,hall,spin,shell,fur3,b}. Hence, in this work, we investigate the influence of a screw dislocation on confinement of a spinless particle to the doubly anharmonic oscillator by searching for analytical solutions to the Schr\"odinger equation. We show that the topology of this defect modifies the spectrum of energy of the doubly anharmonic oscillator.

This paper is structured as follows: in section II, we write the Schr\"odinger equation for the doubly anharmonic oscillator in the presence of a screw dislocation and show that it can be solved analytically. Then, we discuss the influence of the topological defect on the spectrum of energy of the the doubly anharmonic oscillator; in section III, we present our conclusions

\section{Doubly anharmonic oscillator in the presence of a screw dislocation}

In this section, our aim is to obtain the analytical solutions to the Schr\"odinger equation for a particle confined to the doubly anharmonic oscillator in the presence of a screw dislocation. Here, we consider a screw dislocation that corresponds to the distortion of a circular curve into a vertical spiral \cite{put,val}. This kind of topological defect is described by the line element:
\begin{eqnarray}
ds^{2}=dr^{2}+r^{2}d\varphi^{2}+\left(dz+\chi\,d\varphi\right)^{2}, 
\label{1.1}
\end{eqnarray}
where we have used the units $c=\hbar=1$, and the parameter $\chi$ is a constant that characterizes the torsion field (dislocation). In this case, the torsion corresponds to a singularity at the origin \cite{put,fur3,kleinert,kat}. In the general relativity context, Eq. (\ref{1.1}) gives rise to the spatial part of the line element of a spacetime with a space-like dislocation \cite{put,vb}. 

Next, by following Ref. \cite{doub3}, the doubly anharmonic oscillator is given by the scalar potential
\begin{eqnarray}
V\left(r\right)=\omega\,r^{2}+\lambda\,r^{4}+\eta\,r^{6},
\label{1}
\end{eqnarray}
where $\eta>0$. Thereby, the time-independent Schr\"odinger equation for the doubly anharmonic oscillator in the presence of the screw dislocation is
\begin{eqnarray}
\mathcal{E}\psi&=&-\frac{1}{2m}\left(\frac{\partial^{2}}{\partial r^{2}}+\frac{1}{r}\frac{\partial\psi}{\partial r}\right)\psi-\frac{1}{2m\,r^{2}}\left(\frac{\partial}{\partial\varphi}+\chi\frac{\partial}{\partial z}\right)^{2}\psi-\frac{1}{2m}\frac{\partial^{2}\psi}{\partial z^{2}}\nonumber\\
[-2mm]\label{1.2}\\[-2mm]
&+&\omega\,r^{2}\psi+\lambda\,r^{4}\,\psi+\eta\,r^{6}\,\psi.\nonumber
\end{eqnarray}

Based on the cylindrical symmetry of the system, we can write the solution to Eq. (\ref{1.2}) as $\psi\left(r,\,\varphi,\,z\right)=e^{il\varphi+ikz}\,R\left(r\right)$, where $k=\mathrm{const}$ and $l=0,\pm1,\pm2,\pm3\ldots$ are the eigenvalues of the operators $\hat{p}_{z}=-i\partial_{z}$ and $\hat{L}_{z}=-i\partial_{\varphi}$, respectively. In this way, from  Eq. (\ref{1.2}), we obtain a radial equation given in the form:
\begin{eqnarray}
R''+\frac{1}{r}R'-\frac{\left(l-\chi\,k\right)^{2}}{r^{2}}\,R-2m\omega\,r^{2}R-2m\lambda\,r^{4}-2m\eta\,r^{6}R+\left(2m\mathcal{E}-k^{2}\right)R=0.
\label{1.3}
\end{eqnarray}

Let us proceed our discussion by defining
\begin{eqnarray}
\xi=\frac{\left(2m\eta\right)^{1/4}}{\sqrt{2}}\,r^{2},
\label{1.4}
\end{eqnarray}
and then, the radial equation becomes
\begin{eqnarray}
R''+\frac{1}{\xi}\,R'-\frac{\gamma^{2}}{4\xi^{2}}\,R-a\,\xi\,R-\xi^{2}\,R+\frac{b}{\xi}\,R-c\,R=0,
\label{1.5}
\end{eqnarray}
where we have defined the following parameters in Eq. (\ref{1.5}):
\begin{eqnarray}
\gamma&=&l-\chi\,k;\nonumber\\
a&=&\frac{2m\lambda}{\sqrt{2}\,\left(2m\eta\right)^{3/4}};\nonumber
\label{1.6}\\
b&=&\frac{2m\mathcal{E}-k^{2}}{2\sqrt{2}\,\left(2m\eta\right)^{1/4}};\nonumber\\
c&=&\frac{m\omega}{\left(2m\eta\right)^{1/2}}.\nonumber
\end{eqnarray}

By analysing the asymptotic behaviour of Eq. (\ref{1.5}), the solution $R\left(\xi\right)$ is given in the form:  
\begin{eqnarray}
R\left(\xi\right)=e^{-\frac{\xi^{2}}{2}}\,e^{-\frac{a}{2}\,\xi}\,\xi^{\frac{\left|\gamma\right|}{2}}\,H_{B}\left(\left|\gamma\right|,\,a,\,\frac{a^{2}}{4}-c,\,-2b;\,\xi\right),
\label{1.7}
\end{eqnarray}
where $H_{B}\left(\left|\gamma\right|,\,a,\,\frac{a^{2}}{4}-c,\,-2b;\,\xi\right)$ is the biconfluent Heun function \cite{heun}, i.e., it is the solution to the biconfluent Heun differential equation:
\begin{eqnarray}
H_{\mathrm{B}}''+\left[\frac{\left|\gamma\right|+1}{\xi}-a-2\xi\right]H_{\mathrm{B}}'+\left[\frac{a^{2}}{4}-c-2-\left|\gamma\right|-\frac{\left(a\left[\left|\gamma\right|+1\right]-2b\right)}{2\xi}\right]H_{\mathrm{B}}=0.
\label{1.8}
\end{eqnarray}

With the aim of achieving bound state solutions, let us take $H_{\mathrm{B}}\left(\xi\right)=\sum_{k=0}^{\infty}f_{k}\,\xi^{k}$ \cite{abra,arf,griff}, then, from Eq. (\ref{1.8}), we obtain a relation of the coefficients $f_{1}$ to the coefficient $f_{0}$ given by 
\begin{eqnarray}
f_{1}=\frac{a}{2}\,f_{0}-\frac{b}{1+\left|\gamma\right|}\,f_{0}.
\label{1.9}
\end{eqnarray}
Besides, we obtain the following recurrence relation: 
\begin{eqnarray}
f_{k+2}=\frac{a\left(2k+\left|\gamma\right|+3\right)}{2\left(k+2\right)\left(k+2+\left|\gamma\right|\right)}\,f_{k+1}-\frac{\left(\frac{a^{2}}{4}-c-2-\left|\gamma\right|-2k\right)}{\left(k+2\right)\left(k+2+\left|\gamma\right|\right)}\,f_{k}.
\label{1.10}
\end{eqnarray}

Hence, from  Eq. (\ref{1.10}), we have that the biconfluent Heun series terminates when we impose two conditions:
\begin{eqnarray}
f_{n+1}=0;\,\,\,\,\,\,\,\frac{a^{2}}{4}-c-2-\left|\gamma\right|=2n,\,\,\,\left(n=1,2,3,\ldots\right)  
\label{1.11}
\end{eqnarray}
In this way, we obtain a polynomial solution to the function $H_{\mathrm{B}}\left(\xi\right)$. Let us exemplify the analysis of the conditions given in Eq. (\ref{1.11}) by searching for a polynomial of first degree  $\left(n=1\right)$. Therefore, from the condition $\frac{a^{2}}{4}-c-2-\left|\gamma\right|=2n$, we obtain the relation
\begin{eqnarray}
\lambda_{1,\,l}=\sqrt{\frac{\left(2m\eta\right)^{3/2}}{m^{2}}\left(8+2\left|\gamma\right|\right)+4\omega\eta},
\label{1.12}
\end{eqnarray}
which means that the parameter $\lambda$ can be adjusted with the purpose of achieving the polynomial of first degree. For this reason we have labelled $\lambda=\lambda_{n,\,l}$ in Eq. (\ref{1.12}).

Furthermore, from the condition $f_{n+1}=0$, we have for $n=1$ that $f_{n+1}=f_{2}=0$, then, we obtain the following second degree algebraic equation:
\begin{eqnarray}
\left(2m\mathcal{E}_{1,\,l}-k^{2}\right)^{2}&-&\frac{4m\lambda_{1,\,l}\left(2+\left|\gamma\right|\right)}{\sqrt{2m\eta}}\,\left(2m\mathcal{E}_{1,\,l}-k^{2}\right)+\frac{2m\lambda_{1,\,l}^{2}}{\eta}\left(3+\left|\gamma\right|\right)\left(1+\left|\gamma\right|\right)\nonumber\\
&-&16\left(1+\left|\gamma\right|\right)\sqrt{2m\eta}=0,
\label{1.13}
\end{eqnarray}
where $\mathcal{E}_{1,\,l}$ is the parameter associated with the energy of the ground state of the system. By solving Eq. (\ref{1.13}), we obtain
\begin{eqnarray}
\mathcal{E}_{1,\,l}&=&\frac{\left(2+\left|\gamma\right|\right)}{\sqrt{2m\,\eta}}\sqrt{4m^{2}\omega\,\eta+\left(2m\,\eta\right)^{3/2}\left(8+2\left|\gamma\right|\right)}+\frac{k^{2}}{2m}\nonumber\\
[-2mm]\label{1.14}\\[-2mm]
&\pm&\frac{1}{m\left(2m\,\eta\right)^{3/2}}\sqrt{\left(2m\,\eta\right)^{7/2}\left(12+6\left|\gamma\right|\right)+16m^{4}\omega\,\eta^{3}},\nonumber
\end{eqnarray} 
which are the allowed energies of the ground state of the system. Thereby, both conditions given in Eq. (\ref{1.11}) have been satisfied and a polynomial of first degree to $H_{\mathrm{B}}\left(\xi\right)$ has been achieved. In this case, the wave function (\ref{1.7}) becomes
\begin{eqnarray}
R_{1,\,l}\left(\xi\right)=e^{-\frac{\xi^{2}}{2}}\,e^{-\frac{a}{2}\,\xi}\,\xi^{\frac{\left|\gamma\right|}{2}}\,\left(1+\frac{a}{2}\,\xi-\frac{b}{1+\left|\gamma\right|}\,\xi\right).
\label{1.16}
\end{eqnarray}

Note that, by taking $\chi=0$ in Eqs. (\ref{1.12}) and (\ref{1.14}), no effect of the topology of the defect on the doubly anharmonic oscillator exists. In this way, we have the allowed energies of the ground state of the doubly anharmonic oscillator in the absence of defect, i.e., 
\begin{eqnarray}
\bar{\mathcal{E}}_{1,\,l}&=&\frac{\left(2+\left|l\right|\right)}{\sqrt{2m\,\eta}}\sqrt{4m^{2}\omega\,\eta+\left(2m\,\eta\right)^{3/2}\left(8+2\left|l\right|\right)}+\frac{k^{2}}{2m}\nonumber\\
[-2mm]\label{1.15}\\[-2mm]
&\pm&\frac{1}{m\left(2m\,\eta\right)^{3/2}}\sqrt{\left(2m\,\eta\right)^{7/2}\left(12+6\left|l\right|\right)+16m^{4}\omega\,\eta^{3}}.\nonumber
\end{eqnarray}

Hence, by comparing Eq. (\ref{1.15}) to Eq. (\ref{1.14}), we observe that the presence of the screw dislocation modifies the degeneracy of the energy levels of the doubly anharmonic oscillator. Besides, the topological effects of the presence of the screw dislocation determine the possible values of the parameter $\lambda_{n,\,l}$ as we can see in Eq. (\ref{1.12}). Then, for each energy levels $n$ of the system, the allowed values of the parameter $\lambda_{n,\,l}$ of the doubly anharmonic oscillator are established by the quantum numbers $\left\{n,\,l\right\}$ and the parameter related to the screw dislocation in order that a polynomial solution to the function $H_{\mathrm{B}}\left(\xi\right)$ can be achieved.

\section{Conclusions}

We have investigated the influence of a distortion of a circular curve into a vertical spiral, i.e., a screw dislocation, on the doubly anharmonic oscillator. We have shown that the Schr\"odinger equation can be solved analytically. Moreover, we have obtained the exact expressions for the allowed energies of the ground state of the system. We have seen that the presence of the linear topological defect modifies the degeneracy of the energy levels of the doubly anharmonic oscillator. We have also seen that there exists a restriction on the values of the parameter $\lambda$ of the doubly anharmonic oscillator in order to obtain a polynomial solution to $H_{\mathrm{B}}\left(\xi\right)$. These values of the parameter $\lambda$ are determined by the parameter related to the screw dislocation and the quantum numbers. Since we have obtained analytical solutions and the exact expression for the allowed energies of the ground state of the doubly anharmonic oscillator in the presence of a screw dislocation, other points of discussion arise from the present analysis. Two interesting points are the possibility of building coherent states \cite{coh,coh2,coh3,coh4} and displaced Fock states \cite{disp,lbf}. Another topics for discussion are the Aharonov-Bohm effect for bound states \cite{ab,pesk,fur5,bzb} and the thermodynamics properties of quantum systems \cite{therm1,therm2,therm3,therm4,therm5,therm6}.









\acknowledgments{The author would like to thank CNPq for financial support.}

\end{document}